\documentclass[pra,reprint,amsfonts,amsmath,amssymb,superscriptaddress,showpacs]{revtex4-1}

\usepackage{graphicx}
\usepackage{dcolumn}
\usepackage{bm}
\usepackage[bookmarks=true,colorlinks=true,linkcolor=black,citecolor=black,filecolor=black,menucolor=black,urlcolor=black]{hyperref}
\usepackage{amsthm}
\usepackage{mathrsfs}
\usepackage{comment}
\usepackage{color}

\newcommand{\bra}[1]{{\langle #1 |}}

\newcommand{\ket}[1]{{| #1 \rangle}}

\newcommand{\bracket}[1]{{\langle #1 \rangle}}

\DeclareMathOperator{\trace}{Tr}
\newcommand{\Trace}[1]{{\trace[ #1 ]}}
\newcommand{\TRace}[1]{{\trace\left[ #1 \right]}}

\DeclareMathOperator{\Ex}{\mathbb{E}}
\DeclareMathOperator{\Var}{Var}

\DeclareMathOperator{\supp}{supp}

\newcommand{\hilbert}{{\mathcal{H}}}

\newcommand{\RHO}{{\hat\rho}}

\newcommand{\LAMBDA}{{\hat{\bm{\lambda}}}}

\newcommand{\THeta}{{\bm{\theta}}}

\newcommand{\pvm}{{\hat P}}
\newcommand{\POVM}{{\bm{E}}}
\newcommand{\povm}{{\hat E}}
\newcommand{\KRAUS}{{\bm{M}}}
\newcommand{\kraus}{{\hat M}}

\DeclareMathOperator{\rank}{rank}
\newcommand{\trans}{{\mathrm{T}}}
\newcommand{\est}{{\text{est}}}

\theoremstyle{definition}

\newcommand{\rnum}[1]{{\expandafter{\romannumeral #1}}}
\newcommand{\Rnum}[1]{{\uppercase\expandafter{\romannumeral #1}}}

\begin{document}

\preprint{APS/123-QED}

\title{Quantum Estimation Theory of Error and Disturbance in Quantum Measurement}
\author{Yu Watanabe}
\affiliation{%
  Department of Physics, University of Tokyo,
  7-3-1, Hongo, Bunkyo-ku, Tokyo 113-0033, Japan
}%
\author{Masahito Ueda}
\affiliation{%
  Department of Physics, University of Tokyo,
  7-3-1, Hongo, Bunkyo-ku, Tokyo 113-0033, Japan
}%
\affiliation{%
  ERATO Macroscopic Quantum Control Project, JST, 
  2-11-16 Yayoi, Bunkyo-ku, Tokyo 113-8656, Japan
}%

\date{\today}

\begin{abstract}
We formulate the error and disturbance in quantum measurement by invoking quantum estimation theory.
The disturbance formulated here characterizes the non-unitary state change caused by the measurement.
We prove that the product of the error and disturbance is bounded from below by the commutator of the observables.
We also find the attainable bound of the product.
\end{abstract}


\maketitle

\section{Introduction}
\label{sec:intro}

Heisenberg discussed a thought experiment about the position measurement of a particle by the $\gamma$-ray microscope
and found the trade-off relation between the error $\varepsilon(\hat x)$ in the position measurement 
and the disturbance $\eta(\hat p)$ to the momentum caused by the measurement process $\eta(\hat p)$~\cite{bib:indeterminacy}:
\begin{equation}
  \varepsilon(\hat x) \eta(\hat p) \gtrsim \hbar.
\end{equation}
This inequality epitomizes the complementarity of quantum measurements:
we cannot perform the measurement of an observable without causing disturbance to its canonically conjugate observable.
At the inception of quantum mechanics, the Kennard-Robertson inequality~\cite{bib:kennard,bib:robertson}
\begin{equation}
  \Delta X \Delta Y \ge \frac{1}{2}|\bracket{[\hat X, \hat X]}|
\end{equation}
was erroneously interpreted as the mathematical formulation of the trade-off relation of error and disturbance in quantum measurement,
where $\bracket{\hat X} := \Trace{\RHO\hat X}$ is the expectation value of $\hat X$ over the quantum state $\RHO$,
the square bracket denotes the commutator, and $(\Delta X)^2 := \bracket{\hat X^2} - \bracket{\hat X}^2$.
However, $\Delta X$ does not depend on the measurement process.
Thus, the Kennard-Robertson inequality reflects the inherent nature of a quantum system alone, and does not concern any trade-off relation of the error and disturbance in the measurement process.

By performing the measurement we obtain some pieces of the information about the quantum state. 
However, the measurement process causes a non-unitary state change 
and decreases the information on the post-measurement state.
Since the information is conserved under the unitary process,
it can characterize the non-unitary effects of the measurement process.
Therefore, it is expected that there exist the trade-off relations between 
the information obtained by the measurement and the information on the post-measurement state.

Ozawa~\cite{bib:ozawa} discussed the measurement processes and defined the error and disturbance, and derive a trade-off relation.
According to his trade-off relation, it is possible to construct the measurement scheme such that the product of the error and disturbance vanishes.
However, this does not mean that we can obtain information about the observable without dicreasing the information about the canonically conjugate observable on the post-measurement state,
since his definitions of the error and disturbance per se do not always give quantitative information concerning observables.

In this paper, we formulate the complementarity of quantum measurements in terms of the information.
Among several types of information contents in quantum theory, we use the Fisher information which gives precision of the estimated value calculated from the measurement outcomes.
Because the measurement is performed to know the expectation value of an observable $\hat X_1$,
it is natural that the error is measured by the precision of the estimated value of $\bracket{\hat X_1}$.
The non-unitary state change caused by the measurement process
hinders us from estimating the expectation value of the conjugate observable.
Thus the disturbance is characterized by the Fisher information corresponding to the estimation from the outcome of the sequential measurement,

This paper is organized as follows.
In Sec.~\ref{sec:def-error-dist}, we define the measurement error and disturbance by invoking quantum estimation theory.
In Sec.~\ref{sec:trade-off}, we derive trade-off relations between the measurement error and disturbance.
In Sec.~\ref{sec:summary}, we summarize the main results of this paper and discuss some outstanding issues.

\section{Error and Disturbance in Quantum Measurement}
\label{sec:def-error-dist}
\subsection{Measurement Error}
\label{sec:error}
\begin{figure*}[htp]
  \centering
  \includegraphics[width=420pt]{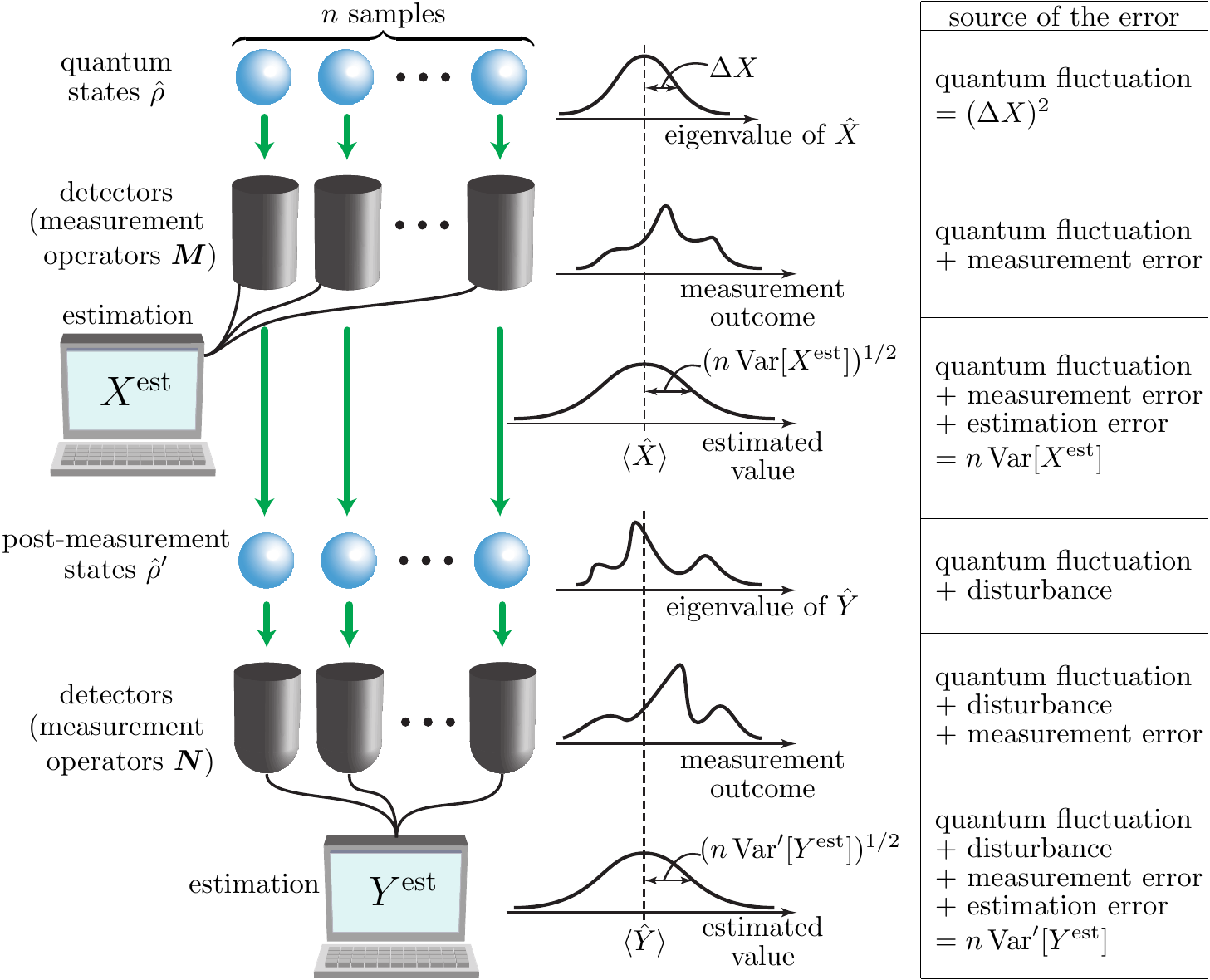}
  \caption{(Color online)
    Measurement process described by measurement operators $\bm{M}$ to know $\bracket{\hat X}$.
    Due to the error in the measurement, the distribution of the measurement outcomes per se does not always provide the quantitative measurement error.
    To retrieve the information contained in the measurement outcomes, it is necessary to estimate $\bracket{\hat X}$ from the measurement outcomes.
    The three types of errors described in the figure contribute to the variance of the estimated value $n\Var[X^\est]$.
    By subtructing the quantum fluctuation and estimation error from $n\Var[X^\est]$, the error inherent in the measurement process is obtained.
    Since the probability distribution on eigenvalues of $\hat Y$ on the post-measurement state $\RHO'$ does not provide quantitative disturbance caused by the measurement $\bm{M}$,
    it is necessary to consider the sequential measurement and estimation process.
    The disturbance is quantified by subtracting the unwanted errors contained in the variance $n\Var'[Y^\est]$.
  }
  \label{fig:flowchart}
\end{figure*}

Suppose we have $n$ independent and identically distributed (i.i.d.) unknown quantum states $\RHO$ on $d$-dimensional Hilbert spaces.
To know the expectation value $\bracket{\hat X}:=\Trace{\RHO\hat X}$ of an observable $\hat X$,
suppose that we perform the same measurement described by meausrement operators $\KRAUS=\{\kraus_{i,a}\}$~\cite{bib:kraus}, where the first index $i$ denotes the measurement outcome.
The probability distribution of the measurement outcomes and the post-measurement state $\RHO'$ are given by
\begin{gather}
  p_i=\TRace{\RHO \sum_a \kraus_{i,a}^\dagger \kraus_{i,a}} = \Trace{\RHO\hat E_i}, \\
  \RHO' = \sum_{i,a} \kraus_{i,a}\RHO \kraus_{i,a}^\dagger,
\end{gather}
where $\bm{E} = \{\hat E_i\}$ is the positive operator-valued measure (POVM) corresponding to $\bm{M}$.
If the measurement is the projection measurement, then the estimated value of $\bracket{\hat X}$ is calculated by 
\begin{equation}
  X^\est = \sum_i \alpha_i \frac{n_i}{n},
\end{equation}
where $\alpha_i$ are the eigenvalues of $\hat X$, and $n_i$ is the number of times that the outcome $i$ is obtained ($n=\sum_i n_i$).
In general, the measurement error affects the outcomes,
and thus the estimation of $\bracket{\hat X}$ is nontrivial.
A reasonable requirement to the estimators is the so-called consistency that for all quantum states $\RHO$ and an arbitrary $\delta > 0$ the estimated value asymptotically converges to $\bracket{\hat X}$:
\begin{equation}
  \lim_{n\rightarrow\infty} \text{Prob}(|X^\est - \bracket{\hat X}| < \delta) = 1. \label{eq:consistent-estimator}
\end{equation}
An example of the consistent estimator is the maximum likelihood estimator.
Since the estimated value is calculated from the measurement outcomes, the estimator of $\bracket{\hat X}$ is a function of $\{n_i\}$: $X^\est = X^\est(\{n_i\})$.
The expectation value and variance of the estimator $X^\est$ are calculated to be
\begin{gather}
  \Ex[X^\est] := \sum_{\{n_i\}} p(\{n_i\}) X^\est(\{n_i\}), \label{eq:ex-x} \\
  \Var[X^\est] := \Ex[(X^\est)^2] - \Ex[X^\est]^2,
\end{gather}
where the summation in \eqref{eq:ex-x} is taken over all sets $\{n_i\}$ that satisfy $n_i \geq 0$ and $\sum_i n_i = n$, and $p(\{n_i\})$ is the probability that each outcome $i$ is obtained $n_i$ times:
\begin{equation}
  p(\{n_i\}) = n!\prod_i \frac{p_i^{n_i}}{n_i!}
\end{equation}
From \eqref{eq:consistent-estimator}, the average of the estimator satisfies
\begin{equation}
  \lim_{n\rightarrow\infty}\Ex[X^\est] = \bracket{\hat X}.
\end{equation}
The variance $\Var[X^\est]$ is caused by three different kinds of errors: the quantum fluctuations, measurement errors and estimation errors (see Fig.~\ref{fig:flowchart}).
The estimation error arises unless we use optimal estimators that minimize $\Var[X^\est]$ such asthe maximum likelihood estimator.

The variance $\Var[X^\est]$ is bounded from below by the Cram\'er-Rao inequality~\cite{bib:cramer}:
\begin{equation}
  \lim_{n\rightarrow\infty} n\Var[X^\est] \geq \bm{x}^\trans J(\KRAUS)^{-1}\bm{x}, \label{eq:cramer-rao}
\end{equation}
where $\trans$ denotes the transpose of the vector, $J(\KRAUS)$ is the Fisher information matrix
\begin{equation}
  [J(\KRAUS)]_{\mu\nu} := \sum_i p_i [\partial_\mu \log p_i][\partial_\nu \log p_i],
\end{equation}
and the column vector $\bm{x}$ is given by
\begin{equation}
  x_\mu = \partial_\mu \bracket{\hat X},
\end{equation}
with $\partial_\mu = \partial/\partial \theta_\mu$, and $\THeta = (\theta_1,\dots,\theta_{d^2-1})$ are real parameters that characterize $\RHO$ such that
any quantum state can be uniquely determined by specifing $\THeta$.
The Fisher information matrix may have $0$ eigenvalues. The right-hand side (RHS) of \eqref{eq:cramer-rao} is calculated to be
\begin{equation}
  \bm{x}^\trans J(\KRAUS)^{-1}\bm{x} =
  \begin{cases}
    \bm{x}^\trans J(\KRAUS)^{+}\bm{x}, &\quad \bm{x}\in\supp[J(\KRAUS)], \\
    +\infty, &\quad \text{otherwise},
  \end{cases}
\end{equation}
where $J(\KRAUS)^{+}$ is the Moore-Penrose pseudoinverse of $J(\KRAUS)$.
The case that the RHS of \eqref{eq:cramer-rao} is infinite means there exists no consistent estimator of $\bracket{\hat X}$.
It occurs, for example, by performing the projection measurement of an observable which does not commute with $\hat X$.
If the RHS of the Cram\'er-Rao inequality \eqref{eq:cramer-rao} is finite, 
there always exist estimators that satisfy the equality of \eqref{eq:cramer-rao} such as the maximum likelihood estimator.
Since such estimators minimize the variance, they are optimal to estimate $\bracket{\hat X}$ from the measurement outcomes,
and $\lim_{n\to\infty}n\Var[X^\est]$ of the optimal estimators, equivalent to the RHS of \eqref{eq:cramer-rao}, does not caused in the estimation process.
Therefore, the RHS of \eqref{eq:cramer-rao} shows the quantum fluctuation and measurement error.

The RHS of \eqref{eq:cramer-rao} is independent of the specification of $\RHO$ by $\THeta$.
Thus, we use the following parameterization.
\begin{equation}
  \RHO = d^{-1}\hat I + \THeta^\trans\LAMBDA = d^{-1}\hat I + \sum_\mu \theta_\mu\hat\lambda_\mu,
\end{equation}
where $\hat I$ is the identity operator, and $\LAMBDA = \{\hat \lambda_1,\dots,\hat\lambda_{d^2-1}\}$ is the generators of the Lie algebra $\mathfrak{su}(d)$.
The generator $\LAMBDA$ satisfy
\begin{equation}
  \hat\lambda_\mu^\dagger = \hat\lambda_\mu, \quad \Trace{\hat\lambda_\mu} = 0, \quad \Trace{\hat\lambda_\mu\hat\lambda_\nu} = \delta_{\mu\nu}.
\end{equation}
In terms of this generator, the observable $\hat X$, and the POVM $\bm{E}$ can be written as
\begin{gather}
  \hat X = x_0 \hat I + \bm{x}^\trans\LAMBDA, \\
  \povm_i = r_i\hat I + \bm{v}_i^\trans\LAMBDA.
\end{gather}
The expectation value $\bracket{\hat X}$ and the probability distribution can be calculated as
\begin{gather}
  \bracket{\hat X} = x_0 + \bm{x}^\trans\THeta, \\
  p_i = r_i + \bm{v}_i^\trans\THeta.
\end{gather}
Then, the RHS of \eqref{eq:cramer-rao} can be calculated to be
\begin{equation}
  \bm{x}^\trans J(\KRAUS)^{-1} \bm{x} = \bm{x}^\trans\left[\sum_i p_i^{-1}\bm{v}_i\bm{v}_i^\trans\right]^{-1} \bm{x}.
\end{equation}

The Fisher information matrix $J(\KRAUS)$ varies with varying $\KRAUS$, but it is bounded from above by the quantum Cram\'er-Rao inequality~\cite{bib:caves:quantum-cramer-rao}:
\begin{equation}
  J(\KRAUS) \leq J_Q, \label{eq:quantum-cramer-rao}
\end{equation}
where $J_Q$ is the quantum Fisher information, that depend only on quantum state $\RHO$.
The quantum Fisher information is a monotone metric on the quantum state space with the coordinate system $\THeta$.
Here, by monotone means that for any quantum operation $\mathcal{O}$ the following inequality is satisfied:
\begin{equation}
  J_Q \geq J_Q',
\end{equation}
where $J_Q'$ is the quantum Fisher information on $\mathcal{O}(\RHO)$.
Although the quantum Fisher information is not uniquely determined,
from the monotonicity condition there exist the minimum and the maximum~\cite{bib:petz}.
The minimum is the symmetric logarithmic derivative (SLD) Fisher information $J_S$~\cite{bib:helstrom:quantum-fisher}.
The SLD Fisher information is a real symmetric matrix, whose $\mu\nu$-element is defined as
\begin{equation}
  [J_S]_{\mu\nu} := \frac{1}{2}\Trace{\RHO \{\hat L_\mu, \hat L_\nu\}}, \label{eq:def-sld-fisher}
\end{equation}
where the curly brackets $\{\ ,\ \}$ denote the anti-commutator, 
and $\hat L_\mu$ is a Hermitian operator called SLD operator defined as the solution to the following operator equation:
\begin{equation}
  \partial_\mu \RHO = \frac{1}{2}\{\RHO, \hat L_\mu\}. \label{eq:def-sld-op}
\end{equation}
The maximum quantum Fisher information is the right logarithmic derivative (RLD) Fisher information $J_R$.
The RLD Fisher information is a Hermitian matrix, whose $\mu\nu$-element is defined as
\begin{equation}
  [J_R]_{\mu\nu} := \Trace{\RHO \hat L'_\nu \hat L'_\mu}, \label{eq:def-rld-fisher}
\end{equation}
where $\hat L'_\mu$ is an operator called RLD operator defined as the solution to the following operator equation:
\begin{equation}
  \partial_\mu \RHO = \RHO\hat L'_\mu. \label{eq:def-rld-op}
\end{equation}
The inverse of the SLD and RLD Fisher information matrices are calculated to be
\begin{gather}
  [J_S^{-1}]_{\mu\nu} = \mathcal{C}_s(\hat\lambda_\mu, \hat\lambda_\nu) := \frac{1}{2}\bracket{\{\hat \lambda_\mu, \hat \lambda_\nu\}} - \bracket{\hat \lambda_\mu}\bracket{\hat \lambda_\nu}, \\
  [J_R^{-1}]_{\mu\nu} = \mathcal{C}(\hat\lambda_\mu, \hat\lambda_\nu) := \bracket{\hat \lambda_\mu \hat \lambda_\nu} - \bracket{\hat \lambda_\mu}\bracket{\hat \lambda_\nu},
\end{gather}
where $\mathcal{C}_s$ and $\mathcal{C}$ are the symmetrized and non-symmetrized correlation functions.
For the observables $\hat X = x_0\hat I + \bm{x}^\trans\LAMBDA$ and $\hat Y = y_0\hat I + \bm{y}^\trans\LAMBDA$,
\begin{gather}
  \bm{x}^\trans J_S^{-1}\bm{x} = \bm{x}^\trans J_R^{-1}\bm{x} = (\Delta X)^2, \label{eq:quantum-fisher-diagonal} \\
  \bm{x}^\trans J_S^{-1}\bm{y} = \mathcal{C}_s(\hat X, \hat Y), \label{eq:sld-off-diagonal} \\
  \bm{x}^\trans J_R^{-1}\bm{y} = \mathcal{C}(\hat X, \hat Y). \label{eq:rld-off-diagonal}
\end{gather}
From \eqref{eq:quantum-cramer-rao} and \eqref{eq:quantum-fisher-diagonal}, the RHS of \eqref{eq:cramer-rao} is bounded from below as
\begin{equation}
  \bm{x}^\trans J(\KRAUS)^{-1}\bm{x} \geq (\Delta X)^2. \label{eq:quantum-cramer-rao-x}
\end{equation}
The equality is achieved if and only if $\bm{M}$ is the projection measurement of $\hat X$,
that is the POVM $\bm{E}$ corresponding to $\bm{M}$ satisfies
\begin{gather}
  \hat E_i\hat E_{i'} = \delta_{ii'}\hat E_i,
  \hat X = \sum_i \alpha_i \hat E_i.
\end{gather}
Since the left-hand side (LHS) shows the quantum fluctuation and measurement error,
and the RHS is the quantum fluctuation, the difference of both sides gives the measurement error.
We define the measurement error as
\begin{equation}
  \varepsilon(\hat X; \KRAUS) := \bm{x}^\trans J(\KRAUS)^{-1}\bm{x} - (\Delta X)^2. \label{eq:error-classical-fisher}
\end{equation}
From \eqref{eq:quantum-cramer-rao-x}, the measurement error $\varepsilon(\hat X; \bm{M})$ is non-negative, 
and vanishes if and only if $\bm{M}$ is the projection measurement of $\hat X$.

Since the Fisher information matrix is defined by the probability distribution of the measurement outcomes, 
the measurement error $\varepsilon(\hat X; \KRAUS)$ is independent of the post-measurement state.
Moreover, if the measurement processes $\bm{M}$ and $\bm{M}'$ satisfy
\begin{equation}
  \bm{M} = \{\hat M_{i,a}\}, \quad \bm{M}' = \{\hat M'_{i,a} = \hat U_{i,a}\hat M_{i,a}\},
\end{equation}
with unitary operators $\hat U_{i,a}$, 
the measurement error $\varepsilon(\hat X; \bm{M})$ and $\varepsilon(\hat X; \bm{M}')$ are equivalent.

\subsection{Disturbance}
\label{sec:disturbance}
Next, we discuss the disturbance caused by the measurement $\KRAUS$.
The disturbance cannot be quantified by the variance of an observable on the post-measurement state.
It is essential to consider another measurement on the post-measurement state and estimation process.
If the disturbance caused by the measurement $\KRAUS$ is small, 
then we can accurately estimate the expectation value of another observable $\hat Y$ from the post-measurement state by performing an appropriate measurement.
If the disturbance causes a drastic state change, then it is hard to estimate $\bracket{\hat Y}$ from the post-measurement state.
Suppose that we perform the measurement $\bm{N} = \{\hat N_{j,b}\}$ on the post-measurement state $\RHO'$.
The probability distribution of the measurement outcomes is given by
\begin{equation}
  q_j = \sum_b\Trace{\RHO'\hat N_{j,b}^\dagger N_{j,b}}.
\end{equation}
The estimated value of $\bracket{\hat Y}$ is calculated from the outcomes of the measurement $\bm{N}$.
The average and the variance of the estimator $Y^\est$ are 
\begin{gather}
  \Ex'[Y^\est] := \sum_{\{n_j\}} q(\{n_j\})Y^\est(\{n_j\}), \label{eq:ex-y}\\
  \Var'[Y^\est] := \Ex'[(Y^\est)^2] - \Ex'[Y^\est]^2,
\end{gather}
where $n_j$ is the number of times that the outcome $j$ is obtained, 
the summation in \eqref{eq:ex-y} is taken over all sets $\{n_j\}$ that satisfy $\sum_j n_j = n$, 
and the probability $q(\{n_j\})$ is 
\begin{equation}
  q(\{n_j\}) = n!\prod_j \frac{q_j^{n_j}}{n_j!}.
\end{equation}
The variance $\Var'[Y^\est]$ is caused by four kinds of errors: the quantum fluctuation on the original quantum state $\RHO$, the disturbance caused by $\KRAUS$, the measurement error in $\bm{N}$, and the estimation error.
The error in the second measurement $\bm{N}$ and estimation error vanish if we perform the optimal measurements and estimations that minimize $\Var'[Y^\est]$.

From the classical and quantum Cram\'er-Rao inequalities, any consistent estimator of $\bracket{\hat Y}$ satisfies
\begin{equation}
  \lim_{n\rightarrow\infty}n\Var'[Y^\est] \geq \bm{y}^\trans J_S'^{-1}\bm{y}, \label{eq:cq-cramer-rao}
\end{equation}
The RHS implies the quantum fluctuation and disturbance caused by $\KRAUS$. 
The SLD Fisher information matrix may have $0$ eigenvalues.
The RHS of \eqref{eq:cq-cramer-rao} is defined by
\begin{equation}
  \bm{y}^\trans J_S'^{-1}\bm{y} = 
  \begin{cases}
    \bm{y}^\trans J_S'^{+}\bm{y} &\quad \bm{y}\in\supp[J_S'] \\
    +\infty &\quad \text{otherwise}.
  \end{cases}
\end{equation}
That the RHS of \eqref{eq:cramer-rao} is infinite means that for any measurement there does not exist consistent estimator $\bracket{\hat Y}$.

Since the SLD Fisher information $J_S$ is the monotone metric, it satisfies $J_S' \leq J_S$.
Thus we obtain
\begin{equation}
  \bm{y}^\trans J_S'^{-1}\bm{y} \geq \bm{y}^\trans J_S^{-1}\bm{y} = (\Delta Y)^2.
\end{equation}
The difference of both sides corresponds to the disturbance caused by $\KRAUS$.
We define the disturbance caused by $\KRAUS$ as 
\begin{equation}
  \eta(\hat Y; \KRAUS) := \bm{y}^\trans J_S'^{-1}\bm{y} - (\Delta Y)^2. \label{eq:def-disturbance}
\end{equation}
From the definitions of the SLD Fisher information matrix \eqref{eq:def-sld-fisher} and the SLD operators \eqref{eq:def-sld-op},
the SLD Fihser information matrix $J_S'$ is invariant under the unitary transformation: $\RHO' \mapsto \hat U\RHO'\hat U^\dagger$.
If the measurement processes $\bm{M}$ and $\bm{M}'$ satisfy
\begin{equation}
  \bm{M} = \{\hat M_{i,a}\}, \quad \bm{M}' = \{\hat M'_{i,a} = \hat U\hat M_{i,a}\},
\end{equation}
the disturbances $\eta(\hat Y; \bm{M})$ and $\eta(\hat Y;\bm{M}')$ are equivalent.
Thus, the definition \eqref{eq:def-disturbance} of the disturbance in terms of the Fisher information can extract the non-unitary effect in the measurement process.

\section{Trade-off between Measurement Error and Disturbance}
\label{sec:trade-off}
\subsection{Inequalities on Error and Disturbance}
To derive the trade-off relations between error and disturbance in quantum measurement,
we show some inequalities satisfied by the error and disturbance.

In Ref~\cite{bib:opt-msmnt-noisy-sys}, it is shown that there exist the measurement $\bm{N}^{\text{opt}}$ such that
\begin{equation}
  \bm{y}^\trans J'(\bm{N}^{\text{opt}})^{-1}\bm{y} = \bm{y}^\trans J_S(\RHO')^{-1}\bm{y}.
\end{equation}
This measurement $\bm{N}^{\text{opt}}$ is the optimal measurement that retrieves the information about $\bracket{\hat Y}$ from the disturbed state $\RHO'$.
The disturbance $\eta(\hat Y; \KRAUS)$ can be written as
\begin{equation}
  \eta(\hat Y; \KRAUS) = \bm{y}^\trans J'(\bm{N}^{\text{opt}})^{-1}\bm{y} - (\Delta Y)^2.
\end{equation}

Performing measurements $\bm{M} = \{\hat M_{i,a}\}$ and $\bm{N}^{\text{opt}}=\{\hat N_{j,b}^{\text{opt}}\}$ sequencially is equivalent to performing the measurement $\bm{A} = \{\hat A_{ij,ab}\}$ whose elements are 
\begin{equation}
  \hat A_{ij,ab} = \hat N_{j,b}^{\text{opt}} \hat M_{i,a}.
\end{equation}
The probability $r_{i,j}$ that the outcome $i$ and $j$ are obtained is 
\begin{equation}
  r_{i,j} = \TRace{\RHO \sum_{a,b}\hat A_{ij,ab}^\dagger \hat A_{ij,ab}}.
\end{equation}
The probability distributions $p_i$ and $q_j$ are calculated to be 
\begin{equation}
  p_i = \sum_j r_{i,j}, \quad q_j = \sum_i r_{i,j}.
\end{equation}
These imply that the mapping from $r_{i,j}$ to $p_i$ and the mapping to $q_j$ are the Markovian mapping.
From the monotonicity of the Fisher information, we obtain
\begin{gather}
  J(\bm{M}) \leq J(\bm{A}), \\
  J'(\bm{N}^{\text{opt}}) \leq J(\bm{A}),
\end{gather}
where $J(\bm{A})$ is calculated to be 
\begin{equation}
  [J(\bm{A})]_{\mu\nu} = \sum_{i,j} r_{i,j} (\partial_\mu \log r_{i,j})(\partial_\nu \log r_{i,j}).
\end{equation}
Therefore, the noise and disturbance in the measurement $\bm{M}$ satisfy
\begin{gather}
  \varepsilon(\hat X; \KRAUS) \geq \bm{x}^\trans J(\bm{A})^{-1}\bm{x} - (\Delta X)^2 = \varepsilon(\hat X; \bm{A}), \label{eq:lower-bound-error} \\
  \eta(\hat Y; \KRAUS) \geq \bm{y}^\trans J(\bm{A})^{-1}\bm{y} - (\Delta Y)^2 = \varepsilon(\hat Y; \bm{A}), \label{eq:lower-bound-disturbance}
\end{gather}
where the equalities are simultaneously satisfied if and only if that the POVM $\POVM$ satisfies
\begin{equation}
  \rank\povm_i = 1
\end{equation}
for all outcomes $i$, and the associated post-measurement state $\RHO_i = p_i^{-1}\sum_a\hat M_{i,a}\RHO\hat M_{i,a}^\dagger$ satisfies
\begin{equation}
  \RHO_i \RHO_{i'} = 0,\ \text{unless}\ i=i'.
\end{equation}

\subsection{Heisenberg Type Trade-off Relation}
In Ref~\cite{bib:uncertainty}, it is proved that any quantum measurement satisfies
\begin{equation}
  \varepsilon(\hat X;\bm{A})\varepsilon(\hat Y; \bm{A}) \geq \frac{1}{4}|\bracket{[\hat X, \hat Y]}|^2. \label{eq:simul-heisenberg}
\end{equation}
From \eqref{eq:lower-bound-error} and \eqref{eq:lower-bound-disturbance}, we obtain that the noise and disturbance in the measurement $\bm{M}$ satisfies
\begin{equation}
  \varepsilon(\hat X;\bm{M})\eta(\hat Y; \bm{M}) \geq \frac{1}{4}|\bracket{[\hat X, \hat Y]}|^2. \label{eq:error-disturbance-heisenberg}
\end{equation}
The inequalities \eqref{eq:simul-heisenberg} and \eqref{eq:error-disturbance-heisenberg} are similar, but their physical meaning are completely different.
The inequality \eqref{eq:simul-heisenberg} is the trade-off relation of the measurement errors of the two observables, 
and implies that we cannot perform the precise measurements of the non-commutable observables simultaneously.
Since the measurement error is independent of the post-measurement state, \eqref{eq:simul-heisenberg} indicates nothing about the disturbance in the measurement process.
The inequality \eqref{eq:error-disturbance-heisenberg} is the trade-off relation between the error and disturbance in the measurement process,
and implies that we cannot retrieve the information about an observable without dicreasing the information on the post-measurement state.
The trade-off relation originally discussed by Heisenberg is rigorously proved by the inequality \eqref{eq:error-disturbance-heisenberg}.

\subsection{Attainable Bound of Error and Disturbance}
In the previous section, we show that the error and disturbance are bounded by the commutation relation of the observables.
However, the equality of \eqref{eq:error-disturbance-heisenberg} cannot be achieved for all quantum states.
For example, if $\RHO=d^{-1}\hat I$, 
\begin{equation}
  \bracket{[\hat X, \hat Y]} = 0
\end{equation}
for any $\hat X$ and $\hat Y$. Thus, the RHS of \eqref{eq:error-disturbance-heisenberg} vanish.
The measurement error vanish if $\bm{M}$ is the projection measurement of $\hat X$, but in this case the disturbance diverges.
The product of the measurement errors of non-commutable observables cannot vanish.
Therefore, there exist a stronger bound for the error and disturbance.
In this section, we derive the attainable bound of the error and disturbance.

In Ref~\cite{bib:uncertainty}, it is proved that any measurement scheme $\bm{A}$ that 
performs two projection measurements probabilistically satisfies the following stronger inequality:
\begin{equation}
  \varepsilon(\hat X;\bm{A})\varepsilon(\hat Y; \bm{A}) \geq (\Delta_Q X)^2(\Delta_Q Y)^2 - C_Q(\hat X, \hat Y)^2. \label{eq:simul-new}
\end{equation}
Here $\Delta_Q$ and $\mathcal{C}_Q$ are defined as follows.
Let $\hilbert_a$ ($a=A,B,\dots$) be the simultaneous irreducible invariant subspace of $\hat X$ and $\hat Y$, and $\pvm_a$ the projection operator on $\hilbert_a$.
We define the probability distribution as $p_a:=\bracket{\pvm_a}$ and the post-measurement state of the projection measurement $\{\pvm_A,\pvm_B,\dots\}$ as $\RHO_a := \pvm_a \RHO \pvm_a / p_a$.
Then, $\Delta_Q$ and $\mathcal{C}_Q$ are defined as
\begin{gather}
  (\Delta_Q X)^2 := \sum_a p_a \left(\Trace{\RHO_a \hat X^2} - \Trace{\RHO_a \hat X}^2\right), \\
  \mathcal{C}_Q(\hat X, \hat Y) := \sum_a p_a \left(\frac{1}{2}\Trace{\RHO_a \{\hat X, \hat Y\}} - \Trace{\RHO_a \hat X}\Trace{\RHO_a \hat Y}\right).
\end{gather}
From the Schwarz inequality,
\begin{align}
  &\left| \mathcal{C}_Q(\hat X, \hat Y) + \frac{1}{2}\bracket{[\hat X, \hat Y]}\right|^2 \notag \\
  &\quad=\left| \sum_a p_a \left( \Trace{\RHO_a \hat X\hat Y} - \Trace{\RHO_a \hat X}\Trace{\RHO_a \hat Y} \right) \right|^2 \notag \\
  &\quad\leq (\Delta_Q X)^2(\Delta_Q)^2
\end{align}
the following inequality can be obtained:
\begin{equation}
  (\Delta_Q X)^2(\Delta_Q Y)^2 - C_Q(\hat X, \hat Y)^2 \geq \frac{1}{4}|\bracket{[\hat X, \hat Y]}|^2. \label{eq:generalized-schrodinger-ineq}
\end{equation}
Therefore, the bound set by \eqref{eq:simul-new} is stronger than that set by \eqref{eq:simul-heisenberg}.
The importance of the inequality \eqref{eq:simul-new} is that for all states and observables there exist measurement processes that achieve the equality of \eqref{eq:simul-new}.
The inequality \eqref{eq:simul-new} is not proved for all measurement process, but numerically vindicated~\cite{bib:uncertainty}.

From \eqref{eq:lower-bound-error} and \eqref{eq:lower-bound-disturbance}, we obtain the tighter bound for the error and disturbance in the measurement $\bm{M}$:
\begin{equation}
  \varepsilon(\hat X;\bm{M})\eta(\hat Y; \bm{M}) \geq (\Delta_Q X)^2(\Delta_Q Y)^2 - C_Q(\hat X, \hat Y)^2. \label{eq:error-disturbance-new}
\end{equation}
From the conditions for the equality of \eqref{eq:simul-new}, \eqref{eq:lower-bound-error} and \eqref{eq:lower-bound-disturbance}, 
the measurement $\bm{M}$ which achieves the equality of \eqref{eq:error-disturbance-new} is obtained as
\begin{equation}
  \hat M_i = 
  \begin{cases}
    c_1 \ket{i}\bra{\psi_i}, & \quad (i=1,\dots,d), \\
    c_2 \ket{i}\bra{\psi'_{i-d}}, & \quad (i=d+1,\dots,2d),
  \end{cases} \label{eq:opt-msmnt}
\end{equation}
where $c_1$ and $c_2$ are positive with $c_1 + c_2 = 1$, $\ket{\psi_i}$ and $\ket{\psi'_i}$ are the eigenstates of observables $\hat Z_1$ and $\hat Z_2$, respectively, and $\ket{i}$'s are orthogonal to each other.
The observables $\hat Z_1$ and $\hat Z_2$ are the linear combination of the $\hat X$ and $\hat Y$:
\begin{gather}
  \hat X = a_1\hat Z_1 + a_2\hat Z_2, \\
  \hat Y = b_1\hat Z_1 + b_2\hat Z_2,
\end{gather}
satisfying the following equation
\begin{equation}
  \bm{a}^\trans
  \begin{pmatrix}
    c_2 & 0 \\
    0 & -c_1
  \end{pmatrix}
  \begin{pmatrix}
    (\Delta_Q Z_1)^2 & \mathcal{C}_Q(\hat Z_1, \hat Z_2) \\
    \mathcal{C}_Q(\hat Z_1, \hat Z_2) & (\Delta_Q Z_2)^2
  \end{pmatrix}
  \begin{pmatrix}
    c_2 & 0 \\
    0 & -c_1
  \end{pmatrix}
  \bm{b} = 0.
\end{equation}

\section{Summary and Discussion}
\label{sec:summary}
By invoking quantum estimation theory, we define the error and disturbance in the quantum measurement.
The error and disturbance are expressed in terms of the Fisher information that gives the precision of the estimation concerning observables.
We prove that the product of the error and disturbance is bounded from below by the commutation relation of the observables.
Moreover, we find the attainable bound.

The measurement scheme \eqref{eq:opt-msmnt} that achieves the bound set by \eqref{eq:error-disturbance-new} requires
that the Hilbert space $\hilbert'$ of the post-measurement state $\RHO'$ satisfies $\dim\hilbert' \geq 2d$.
If the dimension of $\hilbert'$ is less than $2d$, especially the case $\dim\hilbert'=d$, the bound set by \eqref{eq:error-disturbance-new} may not be attainable.
The bound for the case that $\dim\hilbert'=d$ is an outstanding issue.

\begin{acknowledgments}
This work was supported by 
KAKENHI 22340114,
a Grant-in Aid for Scientific Research on Innovation Areas "Topological Quantum Phenomena" (KAKENHI 22103005),
the Global COE Program ``the Physical Sciences Frontier,''
and the Photon Frontier Network Program, from MEXT of Japan.
Y.W. acknowledge support from JSPS (Grant No. 216681).
\end{acknowledgments}

\end{document}